\documentclass[amsmath,amssymb,amsfonts,aps,prb,letter,twocolumn,superscriptaddress,footinbib]{revtex4-2}

\usepackage{graphicx}
\usepackage{bm}
\usepackage{float}
\usepackage{color}

\renewcommand{\Im}{{\rm Im}}
\renewcommand{\Re}{{\rm Re}}
\newcommand{\sgn}{{\rm sgn}}
  
\newcommand{\GammaK}{\Gamma_{\rm K}}
\newcommand{\DeltaK}{\Delta_{\rm K}}
\newcommand{\DeltaT}{\tilde\Delta}
\newcommand{\SigmaT}{\tilde\Sigma}

\newcommand{\AF}{A_{\rm F}}
\newcommand{\AK}{A_{\rm K}}
\newcommand{\GF}{G_{\rm F}}
\newcommand{\SigmaF}{\Sigma_{\rm F}}

\newcommand{\TK}{T_{\rm K}}

\newcommand{\meV}{{\rm meV}}

\begin{document}


\title{Temperature evolution of the Kondo peak beyond Fermi liquid theory}
\author{David Jacob}

\email{david.jacob@ehu.es}
\affiliation{Departamento de Pol\'{i}meros y Materiales Avanzados: F\'{i}sica, Qu\'{i}mica y Tecnolog\'{i}a, Universidad del Pa\'{i}s Vasco UPV/EHU,
  Av. Tolosa 72, E-20018 San Sebasti\'{a}n, Spain}
\affiliation{IKERBASQUE, Basque Foundation for Science, Plaza Euskadi 5, E-48009 Bilbao, Spain}

\date{\today}

\begin{abstract}
  The limitation of Fermi liquid theory to very low energies and temperatures poses
  a fundamental problem for describing the temperature evolution of the Kondo peak.
  Here Fermi liquid theory for the single impurity Anderson model is extended beyond
  the low-energy and low-temperature regime by means of an ansatz for the impurity self-energy
  based on the accurate description of the Kondo peak by the Frota function, the similarity
  between energy and temperature in the second-order self-energy, and by exploiting Fermi liquid conditions.
  Analytic expressions for the temperature dependence of the Kondo peak height and width derived
  from this ansatz are in excellent agreement with numerical renormalization group data
  for temperatures beyond the Kondo temperature.
  The derived expression thus allows to unambiguously determine the intrinsic Kondo peak width
  and Kondo temperature from finite temperature measurements of the Kondo resonance.
\end{abstract}

\maketitle

Scanning tunneling microscopy (STM) has become an important experimental tool for 
studying magnetic atoms and molecules on metallic substrates~\cite{Madhavan:Science:1998,Li:PRL:1998,
  Manoharan:Nature:2000,Nagaoka:PRL:2002,Knorr:PRL:2002,Heinrich:Science:2004,Zhao:Science:2005,
  Wahl:PRL:2007,Iancu:NL:2006,Hirjibehedin:Science:2007,Otte:NPhys:2008,Oberg:NNano:2014,Karan:PRL:2015}.
In these systems the coupling of the atomic or molecular spin to the conduction electrons in the substrate
can give rise to the Kondo effect~\cite{Hewson:book:1997}: the magnetic moment is screened
due to formation of a total spin-singlet state between the atom or molecule and the conduction electrons.
The Kondo effect is signaled by the appearance of a Kondo-Fano resonance in the STM spectra~\cite{Ujsaghy:PRL:2000,
  Schiller:PRB:2000,Madhavan:PRB:2001,Baruselli:PRB:2015,Frank:PRB:2015}.
Therefore observation of the Kondo effect provides proof for magnetism in the uncoupled species~\cite{Li:PRL:2020,Turco:JACSAu:2023}.
Together with the possibility to
detect magnetic excitations via inelastic spin tunneling~\cite{Heinrich:Science:2004,Hirjibehedin:Science:2007,
  Fernandez-Rossier:PRL:2009,Zitko:NJP:2010}, STM spectroscopy (STS) provides an
excellent means for characterizing the magnetic properties of atoms, molecules and nanoclusters.

The Kondo temperature $\TK$ is the energy scale that controls the low-temperature dynamics
of a Kondo system~\cite{Note1}. Importantly, $\TK$ defines a crossover temperature at which
the  system enters the Kondo regime and the Kondo peak starts to emerge.
In STS, $\TK$ is conveniently determined from the halfwidth $\GammaK^0$ of the Kondo peak, since $k\TK\sim\GammaK^0$
(see below for a more precise definition).
However, STM spectra are often measured at temperatures comparable to
$\TK$, where the Kondo peak is strongly broadened.
In order to estimate the intrinsic (i.e. zero temperature) width of the Kondo peak, the
temperature evolution of the Kondo peak width is recorded and extrapolated to zero temperature.
This, however, requires knowledge about the functional form of the low-temperature evolution of the Kondo
peak width~\cite{Note7}.

\begin{figure}
  \includegraphics[width=\linewidth]{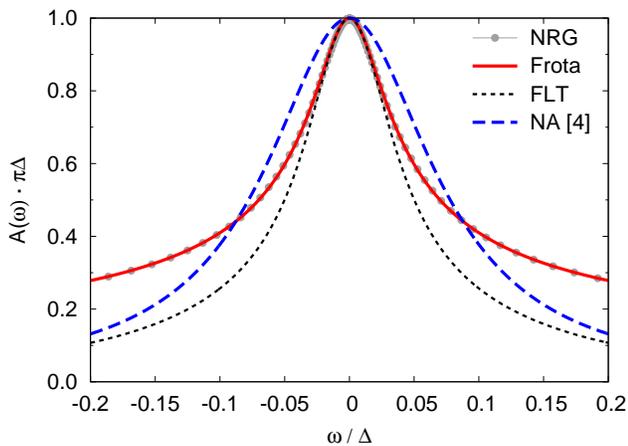}
  \caption{\label{fig:Kondopeaks}
    SFs for the SIAM with $U=-2E_d=10\Delta$ at $T=0$
    calculated by NRG (gray full circles)~\cite{Note4} compared
    to the approximate SFs given by FLT (black dashed line)
    and NA~\cite{Nagaoka:PRL:2002} (blue dashed line).
    Also shown is the Frota lineshape (red solid line)
    fitted to the NRG data with $\DeltaK$ as the fitting
    parameter ($\DeltaK\sim0.0275\cdot\Delta$) and the
    amplitude given by Friedel sum rule, $\AK^0=1/\pi\Delta$.
    The QP weight entering the FLT and NA expressions for
    the SFs is $Z\sim0.055$, obtained from the curvature of the
    NRG SF at the Fermi level.
   }
\end{figure}

Using results from Fermi liquid theory (FLT)~\cite{Nozieres:JLTP:1974,Hewson:PRL:1993,Hewson:book:1997},
Nagaoka {\it et al.} derived a simple expression for the temperature dependence of the Kondo peak's halfwidth~\cite{Note2}:
$\Gamma_{\rm NA}(T)\sim\sqrt{2\,\tilde\Delta^2+(\alpha kT)^2}$ where ${\alpha=\pi}$~\cite{Nagaoka:PRL:2002}
and $\tilde\Delta$ is the renormalized width in FLT (see below) related to the Kondo temperature,
$k\TK\sim\tilde\Delta$.
This equation has been used in a number of papers to estimate the intrinsic width of the Kondo peak
from finite temperature measurements~\cite{Zhao:JCP:2008,Ernst:Nature:2011,Zhang:NComm:2013,
  Khajetoorians:NNano:2015,Gruber:JPCM:2018,Mishra:NNano:2019,Li:PRL:2020,Turco:JACSAu:2023}.
However, in order to fit the experimental data, often
the temperature coefficient $\alpha$ is used as an additional fit parameter, even though according to FLT
$\alpha$ should be exactly $\pi$ in the Kondo regime~\cite{Costi:JPCM:1994,Hewson:book:1997}.
The main problem is the limitation of FLT to very low temperatures and to very
low energies (or bias voltages in STS)~\cite{Chen:PRB:2021,Note8}. For Kondo systems temperature and energy must be 
well below the Kondo temperature and width, effectively one order of magnitude below $\TK$. 
Especially the latter poses a fundamental problem, as it leads to a false estimate of the Kondo peak
width even at zero temperature, see Fig.~\ref{fig:Kondopeaks}.

In order to overcome this problem, in this Letter FLT for Kondo systems
will be extended to a larger range of energies and temperatures. 
Specifically, we focus on the single-impurity Anderson model (SIAM)~\cite{Anderson:PR:1961}:
a single impurity level of energy $E_d$ subject to an on-site Coulomb
repulsion $U$ is coupled to a bath of non-interacting conduction electrons
which gives rise to a constant broadening of the impurity with halfwidth $\Delta$
(wide band limit).
Our starting point is the impurity Green's function (GF) for the particle-hole symmetric SIAM ($E_d=-U/2$)
which according to renormalized perturbation theory (RPT) can be expressed in terms of renormalized quantities
as~\cite{Hewson:PRL:1993}
\begin{equation}
  \label{eq:GF}
  G(\omega) = Z/[\omega+i\DeltaT-\SigmaT(\omega;T)]
\end{equation}
where the chemical potential $\mu$ has been set to zero.
$Z$ is the quasi-particle (QP) weight,
$Z\equiv[1-\partial_\omega\Sigma(0;0)]^{-1}$ where $\Sigma(\omega;T)$ is the electronic
self-energy resulting from the Coulomb repulsion $U$ between
electrons at the impurity site. $\DeltaT\equiv{Z\cdot\Delta}$ is the renormalized halfwidth
of the impurity level, and $\SigmaT(\omega;T)\equiv{Z}(\Sigma(\omega;T)-\Sigma(0;0)-\omega\,\partial_\omega\Sigma(0;0))$
is the \emph{renormalized} self-energy, describing interaction effects between QPs.
Note that at particle-hole symmetry the Hartree contribution to the self-energy $\Sigma(0,0)=U/2$ exactly
cancels $E_d=-U/2$ in the denominator of the GF; also by construction
$\SigmaT(0;0)=\partial_\omega\SigmaT(0;0)=0$. From the GF we can determine the spectral function (SF),
$A(\omega;T)=-\Im\,G(\omega;T)/\pi$, which can be directly related to the $dI/dV$ spectra in
STS~\cite{Jacob:NL:2018,Jacob:JPCM:2018}. 

Perturbation theory to second order~\cite{Yamada:PTP:1975}
yields for the renormalized self-energy ${\SigmaT_2(\omega;T)=-i[\omega^2+(\pi kT)^2]/2\DeltaT}$.
Hence to second order the SF in FLT is given by
$A_{\rm FLT}(\omega;T)=-\tfrac{1}{\pi}\Im\bm(Z/\{\omega+i\DeltaT[1+\tfrac{1}{2}(\omega/\DeltaT)^2+\tfrac{1}{2}(\pi kT/\DeltaT)^2\}\bm)$.
By making a further approximation to the FLT SF~\cite{Note3}
Nagaoka {\it et al.} obtained a Lorentizan form for the SF~\cite{Nagaoka:PRL:2002},
i.e. $A_{\rm NA}(\omega;T)=\tfrac{1}{\pi\Delta}[(\omega/2\DeltaT)^2+1+(\pi kT/2\DeltaT)^2]^{-1}$
with halfwidth given by $\Gamma_{\rm NA}(T)$.

Figure~\ref{fig:Kondopeaks} shows the Kondo peak in the SF computed by numerical
renormalization group (NRG)~\cite{Note4}
compared to the approximate SFs $A_{\rm FLT}(\omega)$ and $A_{\rm NA}(\omega)$ for $T=0$, where
the QP weight $Z$ has been obtained by matching the \emph{curvatures} of
$A_{\rm FLT}$ at $T=0$ 
and the actual Kondo peak in the NRG SF at the Fermi level, i.e., $\partial_\omega^2A_{\rm FLT}(0;0)=\partial_\omega^2A(0;0)$.
In FLT the halfwidth of the actual Kondo peak $\GammaK^0$ is considerably underestimated,
even though the fit with the actual Kondo peak at low energies $\omega\ll\GammaK^0$ is perfect.
In contrast, the Nagaoka approximation (NA) for the SF considerably overestimates the halfwidth.
Additionally, the NA SF does not correctly capture the curvature of the Kondo peak at the Fermi
level either. Thus while the FLT SF yields a proper low-energy and low-temperature description of the Kondo
peak, the NA actually does not.

The underestimate of the Kondo peak width in $A_{\rm FLT}$ is owed to the low-energy nature
of FLT, limiting the validity of the SFs to energies $\omega\ll\GammaK^0$.
The same problem arises for the temperature dependence
which is likewise limited to very low temperatures $T\ll\TK\sim\GammaK^0$. 
In principle this problem could be solved by including higher order terms in the
perturbation expansion. However, very high order terms would be required to achieve a meaningful
extension of the energy and temperature range of FLT. But with growing order the terms also become
increasingly cumbersome for an analytical treatment~\cite{Lesage:PRL:1999,Hewson:JPCM:2001}. 

On the other hand, the Frota function ${\AF(\omega)=\AK^0\cdot\Re\sqrt{i\DeltaK/(\omega+i\DeltaK)}}$~\cite{Frota:PRB:1992}
yields an essentially exact description of the Kondo peak for energies up to several times the halfwidth $\GammaK^0$,
as shown by the red curve in Fig.~\ref{fig:Kondopeaks}.
$\AK^0$ is the amplitude of the Frota function, while $\DeltaK$ determines its halfwidth via
$\GammaK^0=\sqrt{3+\sqrt{12}}\cdot\DeltaK=2.542\cdot\DeltaK$. 
It is now important to realize that the parameters for the Frota function can be
determined \emph{exactly} from FLT since FLT becomes exact in the limit $\omega\rightarrow0,T\rightarrow0$.
First, Friedel sum rule determines the amplitude of the Kondo peak, resulting in
$\AK^0=1/\pi\Delta$. Second, matching the curvatures of the Frota SF and FLT SF at the
Fermi level, $\partial_\omega^2\AF(0)=\partial_\omega^2A_{\rm FLT}(0;0)$, yields
$\DeltaK=\DeltaT/2=Z\cdot\Delta/2$. This is how $Z$ in Fig.~\ref{fig:Kondopeaks} was
determined in practice: instead of taking the second derivative
of the NRG spectral function numerically, which tends to be very noisy,
first the Frota lineshape was determined via the $\DeltaK$ parameter,
and then the QP weight via $Z=2\DeltaK/\Delta$.

Additionally, the finding $\DeltaK=\DeltaT/2$ allows us to establish an \emph{exact}
relation between the Kondo temperature $\TK$ according to Wilson~\cite{Note1} and the intrinsic
width of the Kondo peak $\GammaK^0$. According to FLT $\pi\DeltaT=4k\TK/w$ where $w=0.4128$
is Wilson's number~\cite{Hewson:book:1997}, hence $\DeltaK=2\,k\TK/\pi w\sim1.542\,k\TK$, and therefore:
\begin{equation}
  \GammaK^0 = 2.542\,\DeltaK = \frac{2.542\times2}{\pi\cdot w} \, k\TK \sim 3.92\,k\TK
\end{equation}
The prefactor of $3.92$ is close to the value of $\sim{3.7}$ found numerically
by Zitko and Pruschke from NRG calculations~\cite{Zitko:PRB:2009}.

We next determine the renormalized self-energy $\SigmaT$ that exactly yields the Frota lineshape at $T=0$.
First, we introduce the ``Frota GF'' whose imaginary part yields the Frota spectral function,
$\GF(\omega) \equiv -\tfrac{i}{\Delta} \sqrt{i\DeltaK/(\omega+i\DeltaK)}$,
where $\DeltaK=\DeltaT/2$.
The corresponding renormalized self-energy that yields $\GF(\omega)$
when plugged into (\ref{eq:GF}) is easily determined to be
$\SigmaF(\omega)={\omega + i2\DeltaK \left(1-\sqrt{1-i\omega/\DeltaK } \right)}$.

The crucial step now is to extend the $T=0$ ``Frota self-energy'' $\SigmaF$ 
to finite temperatures.
Inspired by the symmetry in $\omega$ and $\pi kT$ of the second order contribution
to the self-energy $\SigmaT_2\sim{i[\omega^2+(\pi kT)^2]}$, we make the following
Ansatz for the temperature-dependent $\SigmaT$:
\begin{equation}
  \label{eq:SigmaT}
  \SigmaT(\omega;T) = \Re \, \SigmaF(\omega) + i\, \Im\,\SigmaF[\varepsilon(\omega;T)]
\end{equation}
where $\varepsilon(\omega;T)\equiv\sqrt{\omega^2+(\pi kT)^2}$.
Note that the real part of $\SigmaT$ is crucial to recover the Frota lineshape at $T=0$.

The real and imaginary parts of $\SigmaT$ can be written explicitly as real functions:
\begin{eqnarray}
  \label{eq:ReSigmaT}
  \Re\,\SigmaT(\omega;T) &=& \omega - \sigma_\omega\sqrt2\,\DeltaK\,\sqrt{S(\omega/\DeltaK)-1} \\
  \label{eq:ImSigmaT}
  \Im\,\SigmaT(\omega;T) &=&  2\,\DeltaK - \sqrt2\,\DeltaK\,\sqrt{S(\varepsilon/\DeltaK)+1} 
\end{eqnarray}
where $\sigma_\omega\equiv\sgn(\omega)$ is the sign function and $S(x)\equiv\sqrt{1+x^2}$ has been introduced.
The GF can now be written as:
\begin{equation}
  \label{eq:GTilde}
  G(\omega;T) =\frac{\sqrt2/\Delta}{\sigma_\omega\sqrt{S\left(\tfrac{\omega}{\DeltaK}\right)-1}
    +i\sqrt{S\left(\tfrac{\varepsilon(\omega,T)}{\DeltaK}\right)+1}}
\end{equation}
In the limit $T\rightarrow0$ the GF reduces to the Frota form, given by $\GF(\omega)$.
In the following we concentrate on the spectral function~\cite{Note9} given by the
imaginary part of (\ref{eq:GTilde}) which can be written as
\begin{equation}
  \label{eq:ATilde}
  A(\omega;T) = \frac{\sqrt2}{\pi\Delta} \frac{\sqrt{S\left(\tfrac{\varepsilon(\omega,T)}{\DeltaK}\right)+1}}
  {S\left(\tfrac{\omega}{\DeltaK}\right)+S\left(\tfrac{\varepsilon(\omega,T)}{\DeltaK}\right)}
\end{equation}

\begin{figure}
  \includegraphics[width=\linewidth]{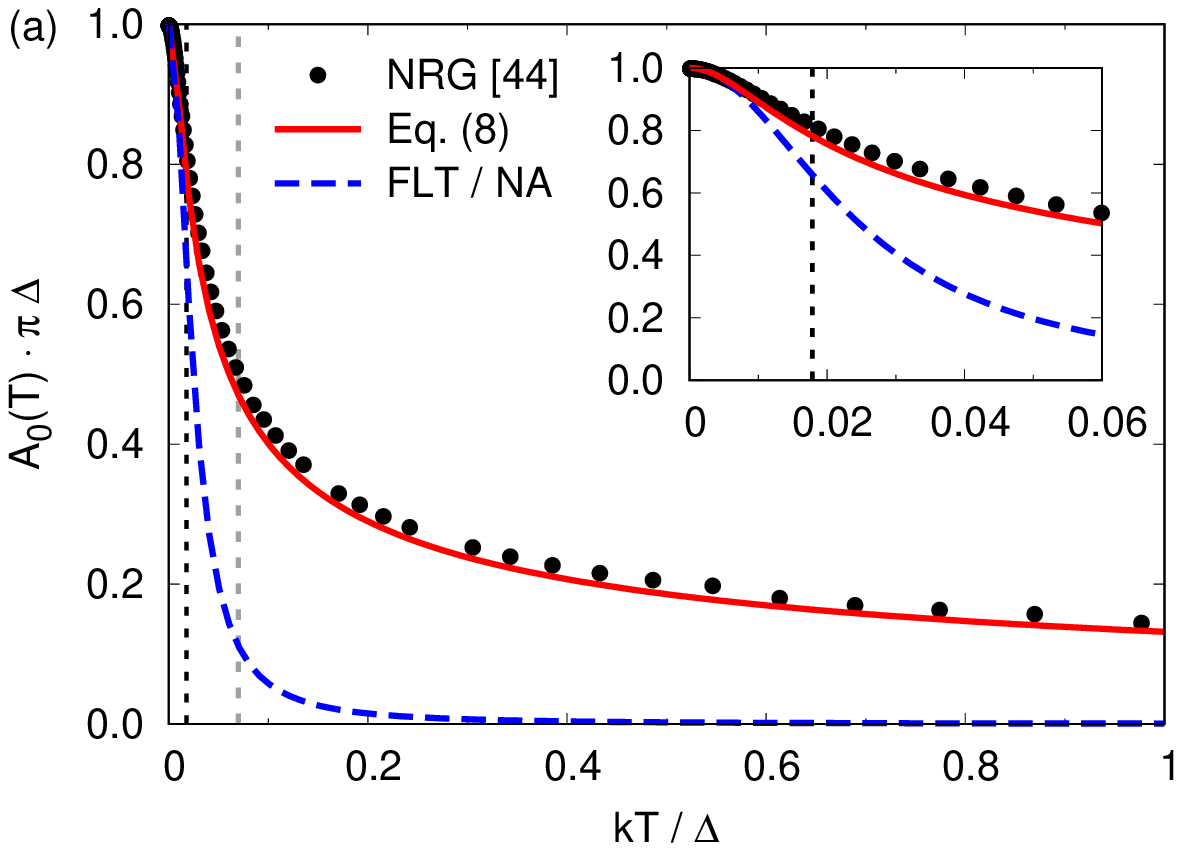}
  \includegraphics[width=\linewidth]{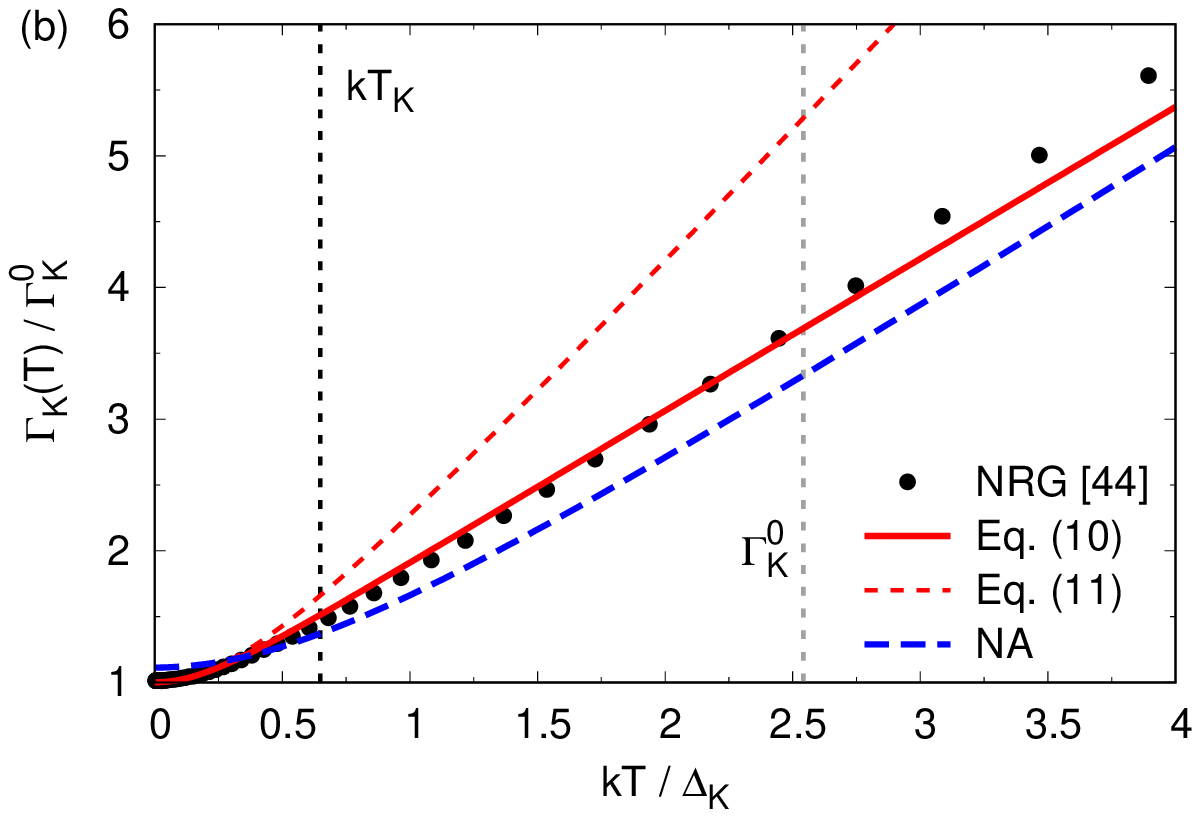}
  \caption{\label{fig:height_and_width}
    Height and halfwidth of Kondo peak as functions of temperature $T$ for the SIAM with $U=-2E_d=10\Delta$.
    (a) Height $A_0(T)$ 
    according to (\ref{eq:height}) (full red line),
    compared to NRG (black circles) \cite{Note4,Osolin:PRB:2013},
    and to FLT / NA (dashed blue line). 
    The inset shows a close-up of the low-temperature region.
    (b) Halfwidth $\GammaK(T)$ according to (\ref{eq:width}) (full red line),
    compared to NRG (black circles)~\cite{Note4,Osolin:PRB:2013},
    and to the NA given by $\Gamma_{\rm NA}(T)$ (blue dashed line)~\cite{Nagaoka:PRL:2002}.
    The thin red dashed line shows the low-temperature approximation (\ref{eq:approx_width}).
    The vertical black and gray dashed lines show $k\TK={\DeltaK/1.542}\sim{0.018\,\Delta}$
    and $\GammaK^0={2.542\,\DeltaK}\sim{0.070\,\Delta}$, respectively.
    The same QP weight as in Fig.~\ref{fig:Kondopeaks}, $Z\sim0.055$, has been used.
  }
\end{figure}

A first test for the validity of the Ansatz (\ref{eq:SigmaT}) for the temperature dependent $\SigmaT$ is to compute the 
temperature dependent height of the Kondo peak, found by evaluating $A$ at $\omega=0$:
\begin{equation}
  \label{eq:height}
  A_0(T)=\frac{1}{\pi\Delta} \sqrt{ \frac{2}{1+\sqrt{1+\left(\tfrac{\pi kT}{\DeltaK}\right)^2} } }
\end{equation}
Figure~\ref{fig:height_and_width}(a) shows the height $A_0(T)$ according to (\ref{eq:height})
compared to NRG data~\cite{Osolin:PRB:2013},
and to the height computed within FLT or NA (both approximations coincide for $\omega=0$).
The agreement between (\ref{eq:height}) and NRG is excellent for temperatures up to $\TK$,
and very good for temperatures up to the bare linewidth, $kT\lesssim\Delta$.
In contrast, in FLT (or NA) the decay of the SF with temperature is far too strong,
leading to a severe underestimate of the Kondo peak height already for temperatures $\sim\TK$.

Next we determine the halfwidth of the Kondo peak $\GammaK$ as a function of temperature, which can be obtained
from the condition $A(\GammaK;T)=\tfrac{1}{2}A_0(T)$. Inserting Eqs.~(\ref{eq:ATilde}) and (\ref{eq:height})
and squaring yields:
\begin{equation}
  \label{eq:HWHM-condition}
  S(\varepsilon/\DeltaK)+1 = \frac{1}{4} \frac{\left[S(\Gamma_{\rm K}/\DeltaK) + S(\varepsilon/\DeltaK)\right]^2}{1+S(\pi kT/\DeltaK)}
\end{equation}
Using the identity $[S(\varepsilon/\DeltaK)]^2=[S(\Gamma_{\rm K}/\DeltaK)]^2+(\pi kT/\DeltaK)^2$
in Eq.~(\ref{eq:HWHM-condition}) would lead to a quartic equation for $S(\Gamma_{\rm K}/\DeltaK)$,
which could in principle be solved analytically, but leads to a very long and cumbersome expression for $S(\Gamma_{\rm K}/\DeltaK)$.
Instead we Taylor expand $S(\varepsilon/\DeltaK)\approx{S(\GammaK/\DeltaK)}+(\pi kT/\DeltaK)^2/2S(\GammaK/\DeltaK)$,
leading to ${[S(\GammaK/\DeltaK)+S(\varepsilon/\DeltaK)]^2}\approx{4S(\GammaK/\DeltaK)^2+2(\pi kT/\DeltaK)^2}$.
This approximation leads to a biquadratic equation for $S(\GammaK/\DeltaK)$ which can be solved easily.
Using $\GammaK=\DeltaK\sqrt{S^2-1}$, we finally obtain the halfwidth of the Kondo peak as a function of
temperature~\cite{Note5}:
\begin{equation}
  \label{eq:width}
  \GammaK(T) = \DeltaK \cdot\sqrt{ a + b\,\sqrt{1+\left(\tfrac{\pi kT}{\DeltaK}\right)^2} + c \, \left(\tfrac{\pi kT}{\DeltaK}\right)^2 }
\end{equation}
where $a\equiv1+\sqrt3\sim2.732$, $b\equiv2+\sqrt3\sim3.732$, and $c\equiv\sqrt3/2\sim0.866$ are constants,
and the Frota width parameter $\DeltaK$ yields the Kondo temperature $\TK=\DeltaK/1.542$ and the intrinsic
halfwidth $\GammaK^0=2.542\,\DeltaK$.

Equation~(\ref{eq:width}) is the central result of this paper. As shown in Fig.~\ref{fig:height_and_width}(b),
it is in excellent agreement with NRG data for temperatures up to $\TK$, and 
is very accurate for temperatures up to $\GammaK^0/k\sim2.542\,\DeltaK/k$ where it starts to deviate
more strongly from NRG. 
In contrast, the temperature evolution of the Kondo peak width in the NA given by
$\Gamma_{\rm NA}(T)$ [Eq.~(8) of Ref.~\onlinecite{Nagaoka:PRL:2002}] yields a poor description of the NRG data in the entire
temperature range. The curvature in the NA in the temperature range $kT\le\DeltaK$ is very
different both from the NRG data and from $\GammaK(T)$ given by Eq.~(\ref{eq:width}).
It also leads to an overestimate of $\sim10\%$ for the intrinsic Kondo peak width in agreement with
Fig.~\ref{fig:Kondopeaks}.

Taylor expansion of the inner square root in (\ref{eq:width}) to second order,
$\sqrt{1+(\pi kT/\DeltaK)^2}\approx1+\tfrac{1}{2}(\pi kT/\DeltaK)^2$, yields
an approximate expression for the halfwidth that resembles the expression found by
Nagaoka {\it et al.}~\cite{Nagaoka:PRL:2002}:
\begin{equation}
  \label{eq:approx_width}
  \GammaK(T) \approx \sqrt{ \left(\GammaK^0\right)^2 + \left(\alpha kT\right)^2 }
\end{equation}
where now $\alpha=\sqrt{1+\sqrt3}\cdot\pi\sim5.193$, different from $\pi$
found by Nagaoka {\it et al.}, but also different from the values found by fitting $\alpha$
in the NA to experimental data for spin-1/2 Kondo systems~\cite{Zhang:NComm:2013,Mishra:NNano:2019,Turco:JACSAu:2023}.
Note that Eq.~(\ref{eq:approx_width}) is only valid in the very low temperature regime $T\ll\TK$, as shown by the red dashed line in
Fig.~\ref{fig:height_and_width}(b), which starts to deviate considerably from the exact result (\ref{eq:width})
for $kT\gtrsim0.3\DeltaK\sim0.5\TK$. However, experimental STS data is usually measured at temperatures comparable to
$\TK$, where the approximation (\ref{eq:approx_width}) is not accurate anymore,
explaining fit values of $\alpha$ different from $5.193$.
Recently, it was also pointed out that simple square root expressions
can in general not capture the correct behavior of Kondo linewidth
both in the low and high-temperature regime~\cite{Note8}. 

\begin{figure}[t]
  \includegraphics[width=\linewidth]{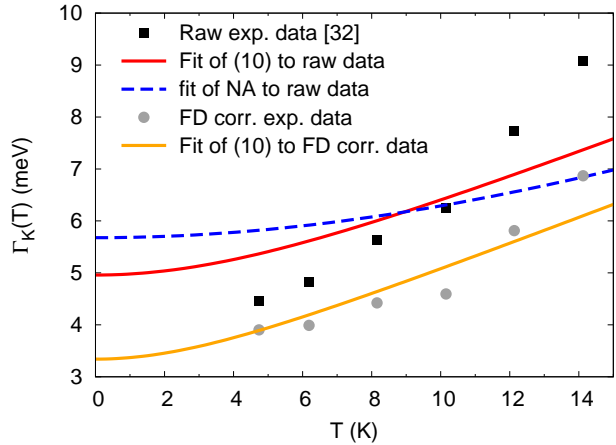}
  \caption{\label{fig:experiment}
    Measured halfwidth of the Kondo peak versus temperature (black solid squares) and fit to
    $\GammaK(T)$ given by (\ref{eq:width}) (full red line)
    for the fused Goblet dimer (data from Ref.~\onlinecite{Mishra:NNano:2019}), resulting in
    $\DeltaK\sim1.95\meV$ corresponding to $\GammaK^0\sim5.0\meV$ and $\TK\sim15$K.
    The blue dashed line shows a fit to the halfwidth $\Gamma_{\rm NA}(T)$ in the NA.
    FD corrected halfwidths~\cite{Note6} are shown as solid gray circles, while the
    full orange line shows a fit of $\GammaK(T)$ given by (\ref{eq:width}) to these, 
    resulting in $\DeltaK\sim1.3\meV$ corresponding to $\GammaK^0\sim3.3$ and $\TK\sim9.9$K.
  }
\end{figure}

Finally, we test how well Eq.~(\ref{eq:width}) can be fitted to existing STS data of a spin-1/2 Kondo system.
Figure~\ref{fig:experiment} shows the temperature evolution of the Kondo halfwidth for the fused Goblet dimer deposited on
Au(111), measured by STS~\cite{Mishra:NNano:2019} (black solid squares) compared to fits of the halfwidth $\GammaK(T)$
given by Eq.~(\ref{eq:width}) (red solid line) and to $\Gamma_{\rm NA}(T)$ in the NA (blue dashed line).
While Eq.~(\ref{eq:width}) performs somewhat better than the NA, it obviously does not fit very well 
the experimental data either, even though the temperatures are well below $\GammaK^0/k\sim37$K~\cite{Mishra:NNano:2019},
where Eq.~(\ref{eq:width}) is expected to be very accurate according to the comparison with NRG, c.f. Fig.~\ref{fig:height_and_width}(b).

A likely explanation for the disagreement is the presence of additional broadening mechanisms
in the STS experiment, often not taken into account in the analysis of the STS data,
as recently discussed by Gruber {\it et al.}~\cite{Gruber:JPCM:2018}.
For example, smearing of the Fermi-Dirac (FD) distribution at the STM tip leads to temperature-dependent
broadening of the $dI/dV$ spectra, described by a convolution of the derivative of the FD distribution and
the spectral function~\cite{Gruber:JPCM:2018,Note8}, which can be evaluated numerically.
The intrinsic halfwidth of the Kondo peak in the underlying spectral function can then
be determined by numerically solving the equation for the effective halfwidth
of the Kondo peak in the $dI/dV$~\cite{Note6}. 
The gray circles in Fig.~\ref{fig:experiment} show the thus FD corrected experimental data.
For the experimental temperature range the effect of FD broadening is considerable (20\% -- 30\%).
As shown by the orange line in Fig.~\ref{fig:experiment}, the FD correction leads to a
considerably better fit of Eq.~(\ref{eq:width}) with the data. Importantly, it leads to a considerably lower estimate of $\GammaK^0$
and $\TK$. Also other broadening mechanisms discussed in Ref.~\onlinecite{Gruber:JPCM:2018} may play a role, and further improve the fit,
when taken into account. The issue of accurately measuring Kondo widths in STS experiments clearly deserves further attention.

In summary, the Fermi liquid description of the Kondo peak has been extended to a larger
energy and temperature range by means of an ansatz for the temperature dependent renormalized
self-energy. The extension beyond Fermi liquid theory is crucial to correctly describe the
width of the Kondo peak at finite temperatures.
Analytic expressions derived from this ansatz for the height and width of the Kondo peak
at finite temperatures show excellent agreement with numerical renormalization group data
up to experimentally relevant temperatures around $\TK$.
The derived expression for the temperature evolution of the Kondo peak width
thus allows to extract the intrinsic Kondo peak width and corresponding Kondo temperature
from finite-temperature STS measurements of Kondo systems.
The discrepancy with published experimental STS data of a spin-1/2 Kondo system~\cite{Mishra:NNano:2019}
can certainly be attributed to the neglect of extrinsic broadening mechanisms in the analysis of the STS data.

\begin{acknowledgments}
I am grateful to Elia Turco, Nils Krane, Pascal Ruffieux, Roman Fasel, Somesh Ganguli, Markus Aapro, Robert Drost, and Peter Liljeroth
for fruitful discussions.
I would also like to thank Rok \v{Z}itko for providing me with the NRG data of Ref.~\onlinecite{Osolin:PRB:2013},
reading of the manuscript and for useful comments. 
I am further grateful to Stefan Kurth and Joaqu\'in Fern\'andez-Rossier who also read the manuscript
and provided useful comments. This work was financially supported by Grant PID2020-112811GB-I00
funded by MCIN/AEI/10.13039/501100011033 and by Grant No. IT1453-22 from the Basque Government.
\end{acknowledgments}

\bibliographystyle{apsrev4-2}
\bibliography{refs,footnotes}

\end{document}


\newcommand{\bra}[1]{\left\langle#1\right|}
\newcommand{\ket}[1]{\left|#1\right\rangle}
\newcommand{\bracket}[2]{\big\langle#1 \bigm| #2\big\rangle}

\newcommand{\Tr}{{\rm Tr}}
\renewcommand{\Im}{{\rm Im}}
\renewcommand{\Re}{{\rm Re}}

\newcommand{\p}{{\prime}}
\newcommand{\pp}{{\prime\prime}}
\newcommand{\ppp}{{\prime\prime\prime}}
\newcommand{\pppp}{{\prime\prime\prime\prime}}

\newcommand{\GammaK}{\Gamma_{\rm K}}
\newcommand{\DeltaK}{\Delta_{\rm K}}
\newcommand{\DeltaT}{\tilde\Delta}
\newcommand{\SigmaT}{\tilde\Sigma}
\newcommand{\epsT}{\tilde\epsilon}
\newcommand{\GT}{\tilde{G}}
\newcommand{\AT}{\tilde{A}}
\newcommand{\AF}{A_{\rm F}}
\newcommand{\AK}{A_{\rm K}}
\newcommand{\GF}{G_{\rm F}}
\newcommand{\SigmaF}{\Sigma_{\rm F}}
\newcommand{\ReF}{F_{\rm R}}
\newcommand{\ImF}{F_{\rm I}}
\newcommand{\ReFb}{\bar{F}_{\rm R}}
\newcommand{\ImFb}{\bar{F}_{\rm I}}
\newcommand{\TK}{T_{\rm K}}

\newcommand{\Seps}{S_{\varepsilon}}
\newcommand{\SGam}{S_{\Gamma}}
\newcommand{\Stau}{S_{\tau}}

\renewcommand\theequation{S\arabic{equation}}
\renewcommand\thefigure{S\arabic{figure}}

\title{Supplemental Material: Accurate temperature evolution of the Kondo peak beyond Fermi liquid theory}

\author{David Jacob}
\email{david.jacob@ehu.es}
\affiliation{Departamento de Pol\'{i}meros y Materiales Avanzados: F\'{i}sica, Qu\'{i}mica y Tecnolog\'{i}a, Universidad del Pa\'{i}s Vasco UPV/EHU,
  Av. Tolosa 72, E-20018 San Sebasti\'{a}n, Spain}
\affiliation{IKERBASQUE, Basque Foundation for Science, Plaza Euskadi 5, E-48009 Bilbao, Spain}

\maketitle

\section{Derivation of Eq.~(10) from Eq.~(9) in main text}

We introduce the abbreviations $\Seps\equiv{S(\varepsilon/\DeltaK)}$, $\SGam\equiv{S(\GammaK/\DeltaK)}$, $\tau\equiv\pi kT/\DeltaK$
and $\Stau\equiv{S(\tau)}$, where $S(x)=\sqrt{1+x^2}$ as defined in the main text after Eq.~(6).
Eq.~(9) of the main text then becomes
\begin{equation}
  \label{eq:hwhm-condition}
  \Seps+1 = \frac{1}{4} \frac{\left(\SGam + \Seps \right)^2}{1+\Stau}
\end{equation}
Using $\Seps^2=\SGam^2+\tau^2$ we can rewrite the numerator of the r.h.s. of (\ref{eq:hwhm-condition}) as
\begin{equation}
   (\SGam+\Seps)^2 = \SGam^2 + 2\,\SGam \, \Seps + \Seps^2 = 2\,\SGam^2 +\tau^2 + 2\,\SGam \, \Seps 
\end{equation}
For $\tau<\SGam$ we can Taylor expand $\Seps$ as $\Seps=\sqrt{\SGam^2+\tau^2}\approx\SGam+\frac{\tau^2}{2\SGam}$
Hence the numerator of the r.h.s. of (\ref{eq:hwhm-condition}) can be approximated by
\begin{equation}
  \label{eq:approx}
  (\SGam+\Seps)^2 \approx 4\SGam^2 + 2\tau^2 \mbox{ for } \tau\ll\SGam
\end{equation}
Eq.~(\ref{eq:hwhm-condition}) then becomes
\begin{equation}
  \sqrt{ \SGam^2 + \tau^2 } \approx \frac{1}{4} \frac{4\SGam^2 + 2\tau^2}{1+\Stau} - 1
\end{equation}
Squaring both sides and reorganizing the terms yields a bi-quadratic equation for $\SGam$:
\begin{equation}
  \left(\SGam^2\right)^2 + p(\tau) \, \SGam^2 + q(\tau) = 0
\end{equation}
where
\begin{equation}
  p(\tau) = -4(1+S_\tau) < 0 \hspace{1ex} \mbox{ and } \hspace{1ex}
  q(\tau) =  -\tfrac{3}{4}\Stau^4 - 3\Stau^3 -\tfrac{1}{2}\Stau^2+ 5\Stau + 4 -\tfrac{3}{4}
\end{equation}
Solving for $\SGam^2$ yields the solution
\begin{equation}
  \label{eq:SGamma2}
  \SGam^2 = -\frac{p(\tau)}{2}  + \sqrt{ \left(\frac{p(\tau)}{2}\right)^2 - q(\tau) } 
\end{equation}
where the positive sign has been chosen in order to guarantee $\SGam^2>0$ for all $\tau$.
The expression under the square root can be simplified to yield:
\begin{equation}
  \left(\frac{p(\tau)}{2}\right)^2 - q(\tau) = \tfrac{3}{4} + 3 \Stau + \tfrac{9}{2} \Stau^2 + 3 \Stau^3 + \tfrac{3}{4} \Stau^4
  = \tfrac{3}{4} \, \left( 1 + \Stau \right)^4
\end{equation}
Plugging this into Eq.~(\ref{eq:SGamma2}) and solving for $\GammaK=\DeltaK\sqrt{\SGam^2-1}$ we obtain the
following equation for the halfwidth of the Kondo peak as a function of the temperature:
\begin{equation}
  \GammaK(T) = \DeltaK \cdot \sqrt{ 1 + 2\, \Stau + \sqrt{\tfrac{3}{4}}\left( 1 + \Stau \right)^2 }
\end{equation}
Using $\Stau=\sqrt{1+\tau^2}=\sqrt{1+\left(\tfrac{\pi kT}{\DeltaK}\right)^2}$ and reorganizing the terms,
we obtain Eq.~(10) of the main text.

\section{Comparison of Eq.~(10) with exact numerical solution of Eq.~(9) in main text}

\begin{figure}[t]
  \includegraphics[width=0.5\linewidth]{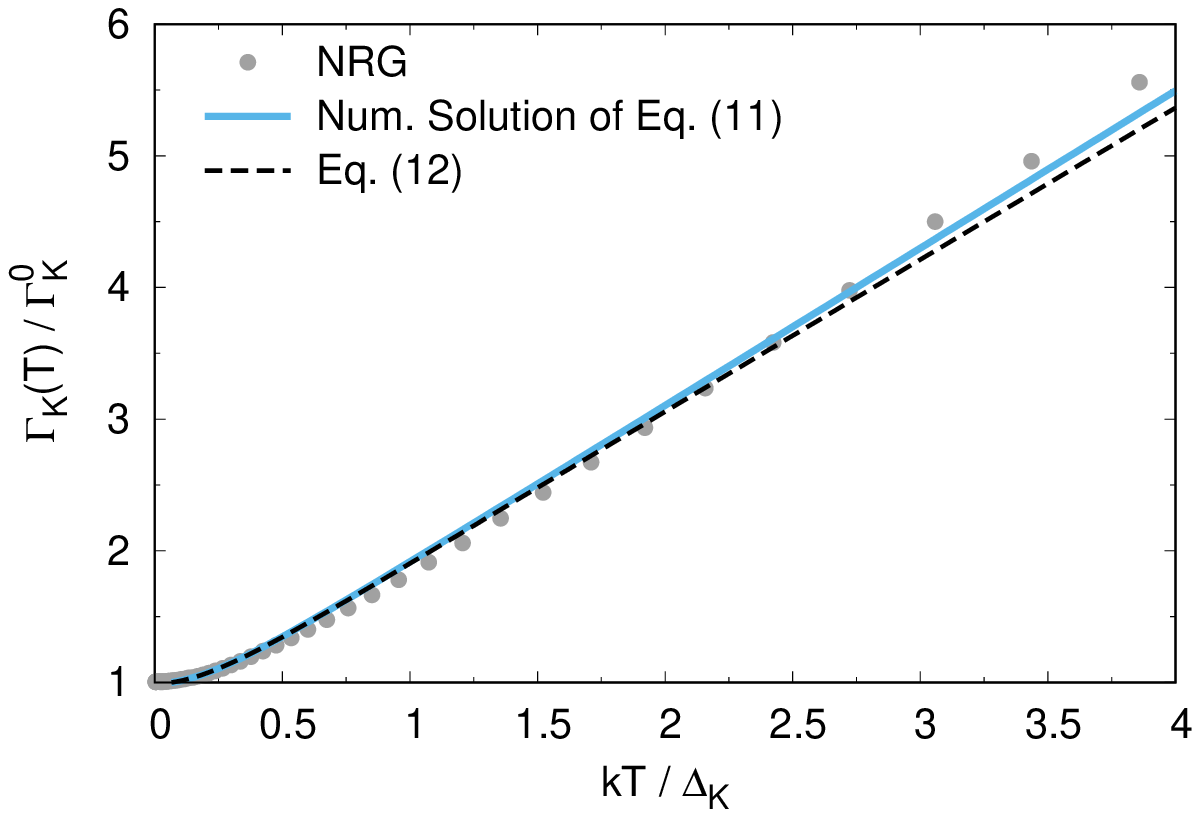}
  \caption{
    \label{fig:numeric} Numeric solution of Eq.(9) compared to the approximate analytic solution (10) in the main text
    and to NRG data for the same Anderson model parameters as in Figs.~1-3 in the main text.
    (NRG data are courtesy of R.~\v{Z}itko and correspond to spectra shown in Fig.~21 of Ref.~\onlinecite{Osolin:PRB:2013}.)
  }
\end{figure}

The approximate analytic solution (10) can be benchmarked by direct numerical solution of  Eq.~(9) or (\ref{eq:hwhm-condition}), respectively.
To this end $\Seps^2=\SGam^2+\tau^2$ is substituted everywhere in (\ref{eq:hwhm-condition}), leading to
\begin{equation}
  \sqrt{\SGam^2+\tau^2} + 1 - \frac{1}{4} \frac{\left(\SGam + \sqrt{\SGam^2+\tau^2} \right)^2}{1+\Stau} = 0
\end{equation}
which can be solved numerically for the root $\SGam$ at each temperature $\tau\equiv\pi kT/\DeltaK$ using, e.g., the bisection method.
The halfwidth is then obtained from $\GammaK=\DeltaK\sqrt{\SGam^2-1}$. Fig.~\ref{fig:numeric} shows the numeric
solution of Eq.~(9) compared to $\GammaK(T)$ given by Eq.~(10) and to the NRG data. Evidently the analytic expression (10) in the main text
yields an excellent approximation to the exact numeric solution of Eq.~(9) in the relevant temperature range $kT\le\GammaK^0=2.542\DeltaK$.

\section{Fermi-Dirac broadening}

At finite temperature $T$ at the STM tip the differential conductance of the current $I$ from the tip to the molecule at the surface
is given by~\cite{Jacob:NL:2018,Gruber:JPCM:2018}:
\begin{equation}
  \label{eq:dIdV}
  \mathcal{G}(V;T,\GammaK)\equiv\frac{d\tilde{I}}{dV} \propto \int d\omega \,  \left[-f^\prime(\omega)\right] \, A(\omega+eV;\GammaK) 
\end{equation}
where $f^\prime(\omega)$ is the derivative of the Fermi-Dirac (FD) distribution $f(\omega)=1/[1+\exp(\beta\omega)]$,
$A(\omega;\GammaK)$ is the spectral function of the molecule, $\beta=1/kT$, $V$ is the applied bias voltage, and $\GammaK$
is the actual width of the Kondo peak in the underlying spectral function $A(\omega;\GammaK)$, that we wish to determine.
The convolution (\ref{eq:dIdV}) can be evaluated numerically.
Approximating the spectral function with a Frota lineshape ${A(\omega;\GammaK)\sim \Re\left[ 1/\sqrt{1+i\omega\cdot2.542/\GammaK }\right]}$,
we can obtain the actual halfwidth $\GammaK$ of the Kondo peak in the spectral function at temperature $T$ as the root of
the function
\begin{equation}
  \label{eq:root}
  g(x) \equiv \mathcal{G}(\tilde{\Gamma}(T)/e;T,x) - \frac{1}{2} \mathcal{G}(0;T,x)
\end{equation}
where $\tilde{\Gamma}(T)$ is the measured halfwidth of the Kondo feature as seen in the $dI/dV$ at temperature $T$
(including FD broadening). The root $g(x)\stackrel{(!)}{=}0$ is found numerically using the bisection algorithm,
and corresponds to the actual halfwidth of the Kondo peak corrected for the FD broadening at temperature $T$, $\GammaK\equiv{x}$.
The FD corrected halfwidths are shown as grey circles for each measured temperature $T$ in Fig.~3 in the main text.

\section{Kramers-Kronig relations}

\begin{figure}
  \includegraphics[width=0.8\linewidth]{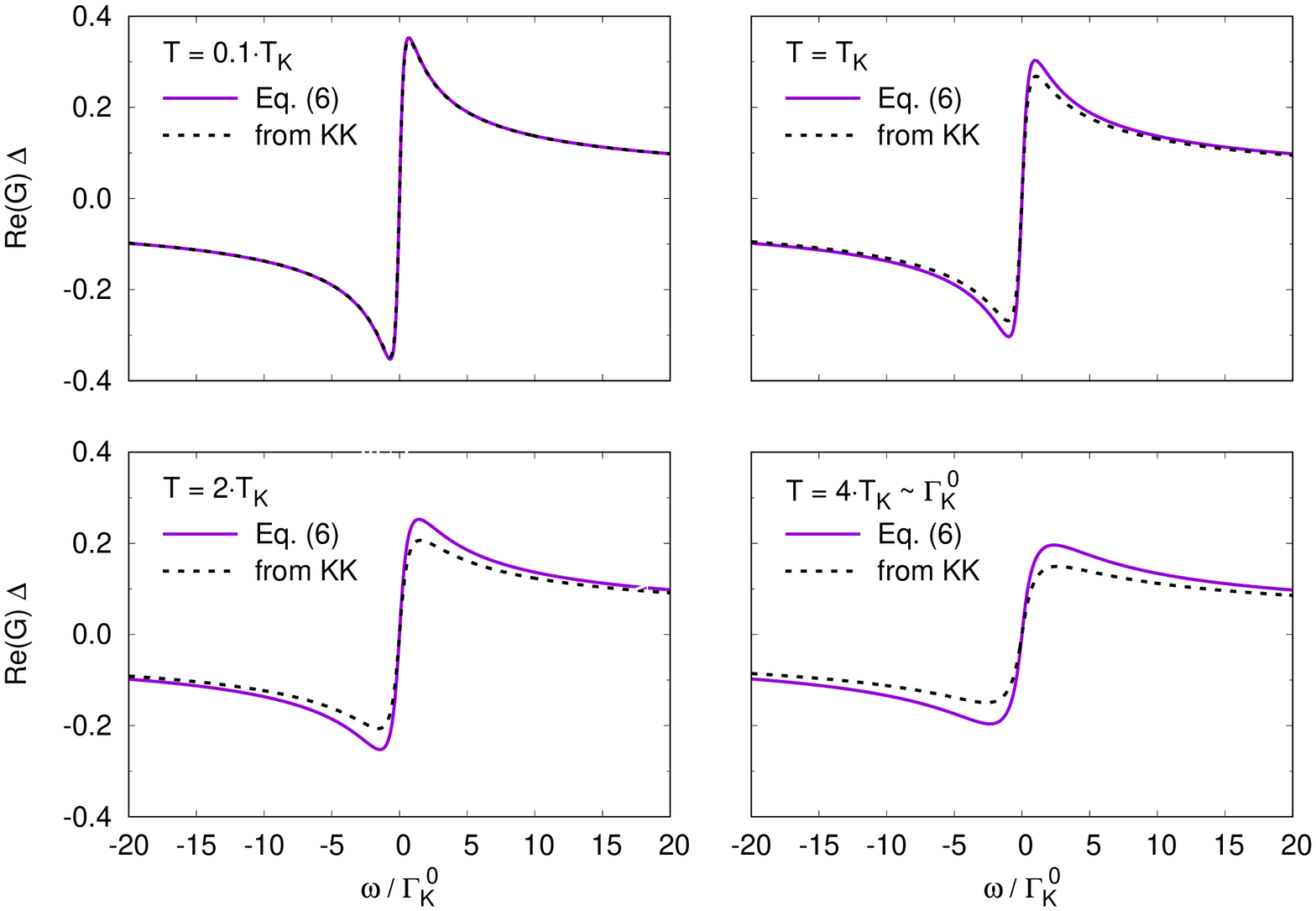}
  \caption{
    \label{fig:KK} Comparison of real part of GF given by Eq.~(6) in main text and real part obtained
    from imaginary part of GF by numerical evaluation of Kramers-Kronig relation (\ref{eq:KK}) for different temperatures.
  }
\end{figure}

As mentioned in footnote [48] of the main text the GF (6) obtained from the Ansatz only approximately fulfills Kramers-Kronig relations for $T>0$.
Fig.~\ref{fig:KK} shows a comparison between the real part of the GF obtained from (6) and the real part obtained from a numerical
evaluation of the Kramers-Kronig relation
\begin{equation}
  \label{eq:KK}
  \Re[G(\omega)] = \frac{1}{\pi} P \, \int \, d\omega^\prime \frac{\Im[G(\omega^\prime+i\eta)]}{\omega^\prime+i\eta-\omega}
\end{equation}
where $i\eta$ is a small imaginary part added in order to avoid the pole at $\omega^\prime=\omega$.
For $T=0$ the GF (6) recovers the Frota form for which Kramers-Kronig is exactly fulfilled.
Also for very small temperatures $T\ll\TK$ the agreement is excellent. For temperatures up to $\TK$ the 
discrepancy is still very small. Only for higher temperatures $T>2\cdot\TK$ the discrepancy becomes appreciable.
In any case for the purpose of the paper the slight inconsistency between real part and imaginary part
of the GF is not of importance, since for the derivation of width and height of the Kondo peak only
the imaginary part or spectral function is of interest.

\bibliographystyle{apsrev4-2}
\bibliography{refs}